\documentstyle[12pt,aps,epsfig]{revtex}

\newcommand{\erf}{\mathop{\rm erf}\nolimits}
\newcommand{\erfc}{\mathop{\rm erfc}\nolimits}
\newcommand{\Bi}{\mathop{\rm Bi}\nolimits}
\newcommand{\sign}{\mathop{\rm sign}\nolimits}

\begin{document}

\title{Third boundary-value problem of the heat conduction equation for a
system with plane-parallel boundaries} \medskip
\author{A.~S.~Usenko\footnote{E-mail address:
usenko@gluk.org}}\medskip
\address{Bogolyubov Institute for Theoretical Physics, Kiev-143, Ukraine
03143}\medskip
\date{\today}

\maketitle
\bigskip

\begin{abstract}
We obtained a new representation of a solution of the heat conduction
\makebox{equation} with boundary condition of the third kind for a layer.
The result is presented as a superposition of fundamental solutions for an
unbounded system with variable coefficients, the explicit form of which is
given. We consider the well-known problem of the evolution of the
temperature field initially uniformly distributed in a layer.  The
distribution of the temperature field is represented in terms of the
obtained functions.
\end{abstract}

\bigskip

\pacs{PACS numbers: 02.30.Jr;44.10.+i} \vfill


\section{Introduction}  \label{Intr.}

Theoretical description of the distribution of heat in bounded solids is
based on the solution of the heat conduction equation with the
corresponding initial and boundary conditions. In the case where the heat
transfer between a heated body occupying the region  $0\leq z\leq L,$
$-\infty < x,y < \infty$  and the external medium with the zero
temperature proceeds by the Newton law, the problem is reduced to the
solution the heat conduction equation \cite{ref.1,ref.2,ref.3} for  $t >
0$

\begin{equation}
 \frac{\partial T(\vec{r},t)}{\partial t} - \kappa\triangle T(\vec{r},t)
  = f(\vec{r},t) \label{Eq. hc}
\end{equation}

\noindent  with the boundary conditions of the third kind

\begin{equation}
 \left. \frac{\partial T(\vec{r},t)}{\partial z} - \lambda_1 T(\vec{r},t)
  = 0 \right |_{z = 0},
 \label{Eq. bc z=0}
\end{equation}

\begin{equation}
 \left. \frac{\partial T(\vec{r},t)}{\partial z} + \lambda_2 T(\vec{r},t)
  = 0 \right |_{z = L}
 \label{Eq. bc z=L}
\end{equation}

\noindent  and the initial condition

\begin{equation}
 T(\vec{r},0) = T_0(\vec{r})
 \label{Eq. ic}
\end{equation}

\noindent  Here,  $\triangle = \partial^2/{\partial x^2} +
\partial^2/{\partial y^2} + \partial^2/{\partial z^2}$  is the Laplace
operator,  $\vec{r} = (x,y,z),$  $\kappa$  is the thermal diffusivity,
$\lambda_1$  and  $\lambda_2$  are the relative heat transfer coefficients
between the body and the external medium at  $z = 0$  and  $z = L,$
respectively, and  $f(\vec{r},t)$  is the density of thermal sources
normalized to  $c\rho,$  where  $c$  and $\rho$  are, respectively, the
specific heat capacity and the density of the substance.

The methods for solving this boundary-value problem are well-known and
widely described in the literature both for the theory of heat conduction
\cite{ref.1,ref.2,ref.3} and for equations of mathematical physics
\cite{ref.4,ref.5,ref.6,ref.7}.  It is worth noting that a solution of
this problem is presented as a series in which the summation is carried
out over an infinite set of eigenvalues  $\alpha$,  where  $\alpha$  are
positive roots of the transcendental equation

\begin{equation}
 \tan \alpha L = \alpha\frac{\lambda_1+\lambda_2}{\alpha^2-\lambda_1 \lambda_2},
 \label{Eq. trans}
\end{equation}

\noindent  which, in the general case, is solved numerically. In
particular cases where the temperature of the boundary surfaces is equal
to zero ($\lambda = \infty$,  the first boundary-value problem) or the
heat flow through the boundaries is absent ($\lambda = 0,$  the second
boundary-value problem), the roots of Eq.~(\ref{Eq. trans}) are known.  In
these cases, the solution of the heat conduction equation is represented
in terms of normal distributions created by instantaneous heat sources
situated at the corresponding points of the space. This enables one,
without recourse to numerical calculations, to carry out analytical
estimations of solutions of the heat conduction equation for the first and
second boundary-value problems for various special cases (thick layer,
near the boundary surfaces and at the center of the layer, time
temperature asymptotics, etc.).  For arbitrary values of  $\lambda,$
Eq.~(\ref{Eq. trans}) was numerically investigated in detail and the
results are presented in the form of tables and plots (see, e.g.,
\cite{ref.3}).

As far as we know, for any  $\lambda > 0$,  in the literature, there is no
representation of a solution of problem (\ref{Eq. hc})--(\ref{Eq. bc z=L})
in terms of fundamental solutions for an unbounded system similar to those
for the first and second boundary-value problems for which it is not
necessary to find eigenvalues  $\alpha$  of the transcendental equation
(\ref{Eq. trans}).  The possibility of this representation for arbitrary
$\alpha$  is shown in the second section of the paper for Green's
function. Unlike the known solutions of boundary-value problems of the
first and second kinds, in this case, the fundamental solutions are in the
sum with the corresponding weights that are functions of coordinates and
time.

As an example, in the third section, we use the obtained Green function
for studying the evolution of an initially uniform temperature field in a
layer.


\section{Green's function of the third boundary-value problem} \label{Green}

We consider the homogeneous heat conduction equation (\ref{Eq. hc}) with
boundary conditions (\ref{Eq. bc z=0})--(\ref{Eq. bc z=L}) for $\lambda_1
= \lambda_2 \equiv \lambda$  and the initial condition

\begin{equation}
 T(\vec{r},0) \equiv T_0(\vec{r}) = \delta(\vec{r}-\vec{r}{\,}^\prime),
 \label{Eq. ic sl}
\end{equation}

\noindent  which corresponds to a point instantaneous heat source situated
at the point  $\vec{r}{\,}^\prime = (x^\prime,y^\prime,z^\prime),$  where
$0 \leq z^\prime \leq L,$  and find a solution continuous in the region $0
\leq z \leq L,$  $-\infty < x,y < \infty$  for  $t > 0.$  Note that in the
general case of different  $\lambda_1$  and  $\lambda_2$, the expression
for Green's function for the third boundary-value problem (\ref{Eq.
hc})--(\ref{Eq. bc z=L}), (\ref{Eq. ic sl}) expressed in terms of
fundamental solutions for an unbounded system was obtained in
\cite{ref.8}, where the Brownian motion of particles in a layer was
investigated for various values of adsorption coefficients of particles by
the boundary surfaces.  We also noted there that the specific case
$\lambda_1 = \lambda_2$  should be considered in its own right because in
going from the general expressions presented in \cite{ref.8} to the case
$\lambda_1 = \lambda_2,$  we have to sum expressions each term of which is
divergent.

By using methods of the operational calculus, we reduce Eq.~(\ref{Eq. hc})
with initial condition (\ref{Eq. ic sl}) to an ordinary differential
equation of the second order, the solution of which  $T(\vec{r},t) \equiv
G(\vec{r},\vec{r}{\,}^\prime,t)$  (Green's function of the heat conduction
equation) can be written as follows:

\begin{eqnarray}
 G(\vec{r},\vec{r}{\,}^\prime,t) &=& \frac{\exp\left(\frac{-R^2_\perp}
 {4\kappa t}\right)}{8\pi^2\kappa t} \sum_{n=-\infty}^{\infty}
 \int\limits_{-\infty}^{\infty}dk_z\, \left\{\exp(-k^2_z\kappa t
 + ik_zZ_-(n))
 \left(\frac{k_z-i\lambda}{k_z+i\lambda}\right)^{2n}\right.
 \nonumber\\
 &+& \left. \exp\left (-k^2_z\kappa t + ik_zZ_+(n)\right )\left
 (\frac{k_z-i\lambda}{k_z+i\lambda}\right)^{2n+1}\right\},
 \label{Eq. G int}
\end{eqnarray}

\noindent  where  $\vec{R}_\perp =
\vec{r}_\perp-\vec{r}{\,}^\prime_\perp,$ $\vec{r}_\perp = (x,y,0),$
$\vec{r}{\,}^\prime_\perp = (x^\prime,y^\prime,0),$ $Z_\pm(n) =
Z_\pm+2nL,$  where  $n = 0,\pm1,\pm2, \ldots,$  $Z_\pm = z \pm z^\prime.$

Then we use the integral representation

\begin{equation}
 \exp(-k^2_z\kappa t) = \frac{1}{(4\pi\kappa t)^{1/2}}\int \limits_
 {-\infty}^{\infty}d \xi\, \exp\left(-\frac{\xi^2}{4\kappa t} + ik_z\xi
 \right),  \label{Eq.8}
\end{equation}

\noindent  change the order of integration, and take integrals over  $k_z$
and  $\xi.$  We present the final result in the following form:

\begin{eqnarray}
 G(\vec{r},\vec{r}{\,}^\prime,t) &=& \frac{\exp\left(\frac{-R^2_\perp}
 {4\kappa t}\right)}{(4\pi\kappa t)^{3/2}} \sum_{n=-\infty}^{\infty}
 \left \{P^{(-)}_n(Z_-(n),\lambda)\exp\left(-\frac{Z^2_-(n)}
 {4\kappa t}\right) \right.
 \nonumber\\
 &+& \left. P^{(+)}_n(Z_+(n),\lambda) \exp \left
 (-\frac{Z^2_+(n)}{4\kappa t}\right)\right\},
 \label{Eq. G main}
\end{eqnarray}

\noindent  where

\begin{eqnarray}
 P^{(-)}_n(z,\lambda) &=& 1 -\left(1-\delta_{n0}\right)
 \sum_{k=0}^{2|n|-1}C^{k+1}_{2|n|}\beta_k(|z|,\lambda),
 \qquad  n = 0,\pm1,\pm2,\ldots,
 \nonumber\\
 P^{(+)}_n(z,\lambda) &=& 1 - \sum_{k=0}^{2n}C^{k+1}_{2n+1}
 \beta_k(z, \lambda),  \qquad  n \ge 0,
 \nonumber\\
 P^{(+)}_n(z,\lambda) &=& 1 - \sum_{k=0}^{2|n|-2}C^{k+1}_{2|n|-1}
 \beta_k(|z|,\lambda),  \quad  n < 0,
 \label{Eq. P n}
\end{eqnarray}

\noindent  $C^n_m$  is the binomial coefficient,  $\delta_{n0}$  is the
Kronecker symbol,  $\tilde{\lambda} = \lambda(\kappa t)^{1/2}$,

\begin{eqnarray}
 \beta_k(z,\lambda) &=& \pi^{1/2}(2\tilde{\lambda})^{k+1}
 F_k\left(\frac{z}{2(\kappa t)^{1/2}} +\tilde{\lambda}\right),
 \qquad  k = 0, 1,2,\ldots,\\
 F_k(x) &=& \frac{1}{k!} \frac{d^k}{dx^k} \left(\exp(x^2)\erfc\,x\right),
 \qquad  k = 0, 1,2,\ldots,\\
 \erfc\, x &=& \frac{2}{\pi^{1/2}} \int \limits_x^{\infty}dy\,\exp(-y^2).
\end{eqnarray}

By the direct substitution of solution (9) into Eqs.~(1)--(3), we can make
sure that this representation of Green's function satisfies both the
homogeneous equation (1) and boundary conditions (2) and (3) (for details,
see Appendix).  To carry out calculations one should use the following
recurrence relationship

\begin{equation}
 F_k(x) = \frac{2}{k}\left[F_{k-2}(x) +x F_{k-1}(x)\right],
 \qquad  k \ge 2
 \label{Eq. F}
\end{equation}

\noindent  for the function  $F_k(x).$  To obtain relation
(\ref{Eq. F}), we use the known recurrence relationships for
multiply probability integrals \cite{ref.9} and representation
(12) for the function $F_k(x).$ As the subscript  $k$  of the
function  $F_k(x)$  increases, the function $F_k(x)$  decreases
and, for all values of the argument  $x,$  remains either positive
(for even  $k$  and  $k = 0$) or negative (for odd $k$). The
behavior of the first six functions  $F_k(x)$  is shown in Fig.~1
[$k = 0,1,2,3,4,5,$  the number near a curve corresponds to the
subscript $k$ of the function  $F_k(x)$].

\begin{figure}
 \centering{\epsfig{figure=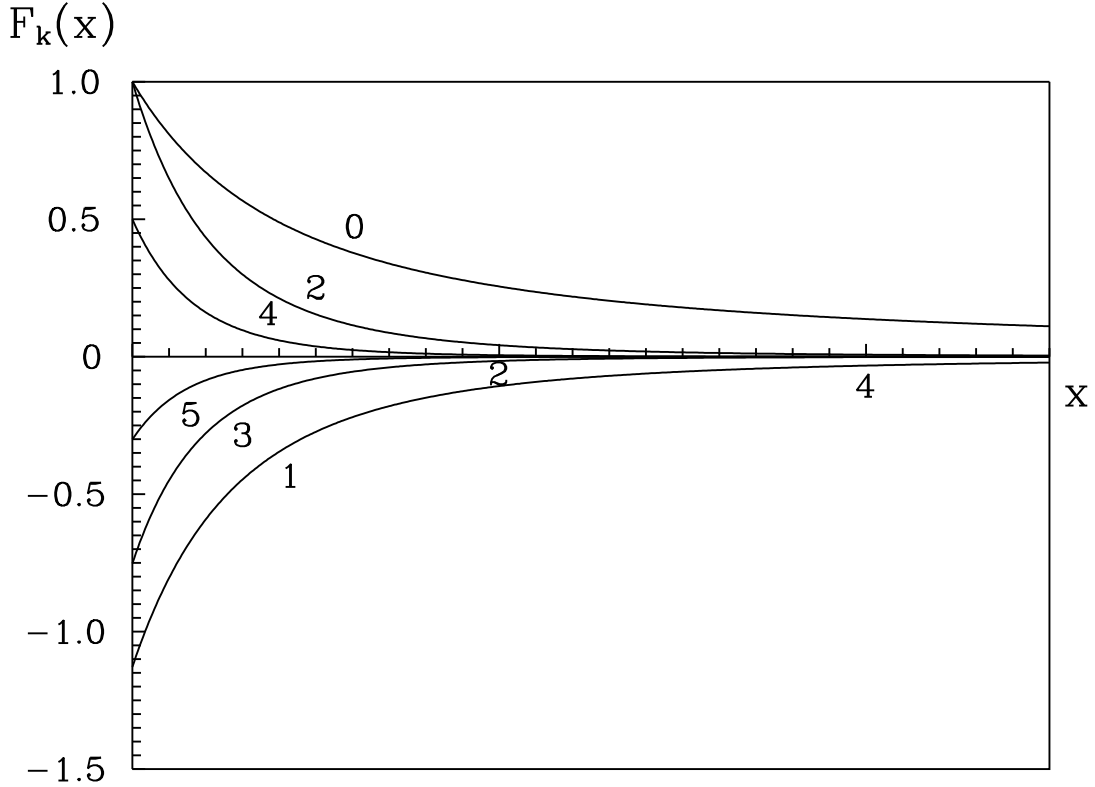}}
 \bigskip
 \caption[]{Function  $F_k(x):$  $ k = 0$ (curve 0), 1 (curve 1), 2
 (curve 2), 3 (curve 3), 4 (curve 4), and 5 (curve 5).}
 \label{fig._1}
\end{figure}

\bigskip

In the limiting case  $\lambda \to \infty,$ with regard for
properties of binomial coefficients, it follows from Eq.~(\ref{Eq.
P n}) that all quantities  $P_n^{(-)}$  and $P_n^{(+)}$  are the
same and constant

\begin{equation}
 \lim\limits_{\lambda \to \infty} P_n^{(\pm)}(Z_{\pm}(n),\lambda) = \pm 1
 \label{Eq. lim}
\end{equation}

\noindent  and Eq.~(\ref{Eq. G main}) is reduced to the well-known
solution of the first boundary-value problem for Green's function in a
layer \cite{ref.1,ref.10}

\begin{eqnarray}
 G(\vec{r},\vec{r}{\,}^\prime,t) &=& \frac{\exp \left(\frac{-R^2_\perp}
 {4\kappa t}\right)}{(4\pi\kappa t)^{3/2}} \sum_{n=-\infty}^{\infty}
 \left\{\exp\left(-\frac{Z^2_-(n)}{4\kappa t}\right) - \exp\left
 (-\frac{Z^2_+(n)}{4\kappa t}\right)\right\},
 \label{Eq. G dif}
\end{eqnarray}

As  $\lambda \to 0,$ [$\lambda(\kappa t)/|Z_{\pm}| \ll 1$], in view of
Eqs.~(10) and (11), we have

\begin{equation}
 \beta_k(|Z_\pm(n)|,\lambda) \approx
   \left \{\begin{array}{ll}
   2\tilde{\lambda}\pi^{1/2} F_0\left(\frac{|Z_\pm (n)|}
   {2(\kappa t)^{1/2}} \right),   \qquad & k=0,\\
   o(\tilde{\lambda}^k),   \qquad & k=1,2,\ldots\\
   \end{array}  \right.
  \label{Eq. 17}
\end{equation}

\noindent  and

\begin{eqnarray}
 P_n^{(-)}(Z_-(n),\lambda) &\approx& 1-4|n|\tilde{\lambda}\pi^{1/2}
 F_0\left(\frac{|Z_-(n)|}{2(\kappa t)^{1/2}} \right),
 \qquad  n=0,\pm1,\pm2,\ldots
 \nonumber\\
 P_n^{(+)}(Z_+(n),\lambda) &\approx& 1-2(2n+1)\tilde{\lambda}\pi^{1/2}
 F_0\left(\frac{Z_+(n)}{2(\kappa t)^{1/2}} \right), \qquad  n \ge 0,
 \nonumber\\
 P_n^{(+)}(Z_+(n),\lambda) &\approx& 1-2(2|n|-1)\tilde{\lambda}\pi^{1/2}
 F_0\left(\frac{|Z_+(n)|}{2(\kappa t)^{1/2}} \right), \qquad  n < 0,
 \label{Eq. P ap d}
\end{eqnarray}

\noindent  whence, for  $\lambda = 0$  $\left[P_n^{(\pm)}(Z_\pm(n),0) =
1\right],$  it follows the well-known solution of the second
boundary-value problem \cite{ref.1,ref.10}

\begin{eqnarray}
 G(\vec{r},\vec{r}{\,}^\prime,t) &=& \frac{\exp \left(\frac{-R^2_\perp}
 {4\kappa t}\right)}{(4\pi\kappa t)^{3/2}} \sum_{n=-\infty}^{\infty}
 \left\{\exp\left(-\frac{Z^2_-(n)}{4\kappa t}\right) + \exp\left
 (-\frac{Z^2_+(n)}{4\kappa t}\right)\right\}
 \label{Eq. G sp}
\end{eqnarray}

In the limiting case of an infinitely thick layer  $\left(L \to \infty
\right),$  we have

\begin{eqnarray}
 P_n^{(-)}(Z_-(n),\lambda) &=& \delta_{n0},
 \nonumber\\
 P_n^{(+)}(Z_+(n),\lambda) &=& \delta_{n0}\left [1-\beta_0(Z_+,\lambda)\right]
 \label{Eq. P inf}
\end{eqnarray}

\noindent and expression~(\ref{Eq. G main}) is reduced to the known
solution of the third boundary-value problem for a half space ($z \ge 0,$
$-\infty < x, y < \infty$) \cite{ref.3,ref.7,ref.10}

\begin{eqnarray}
 G(\vec{r},\vec{r}{\,}^\prime,t) &=& \frac{\exp \left(\frac{-R^2_\perp}
 {4\kappa t}\right)}{(4\pi\kappa t)^{3/2}}
 \left\{\exp\left(-\frac{Z^2_-}{4\kappa t}\right) + \exp\left
 (-\frac{Z^2_+}{4\kappa t}\right)\right.
 \nonumber\\
 &-& \left. 2\tilde{\lambda}\pi^{1/2}\exp({\tilde{\lambda}}^2+\lambda Z_+)
 \erfc\left(\frac{Z_+}{2(\kappa t)^{1/2}} + \tilde{\lambda}\right)\right\}.
 \label{Eq. G inf}
\end{eqnarray}


\section{Evolution of a uniform temperature distribution}  \label{Temperature}

The obtained expression (\ref{Eq. G main}) for Green's function can be
used for the solution of various boundary-value problems for a layer. As
an example, we consider the simplest well-known problem on the evolution
of the temperature in an initially uniformly heated layer
$\left[T_0(\vec{r}) = T_0\right]$  with boundary conditions (2) and (3)
for  $\lambda_1=\lambda_2 \equiv \lambda.$  The required distribution
$T(\vec{r},t)$  is defined by the integral

\begin{equation}
 T(\vec{r},t) = T_0 \int\limits_{-\infty}^{\infty}\!\! dx^\prime
 \int\limits_{-\infty}^{\infty}\!\! dy^\prime
 \int\limits_0^{\infty}\!\! dz^\prime\,G(\vec{r},\vec{r}{\,}^\prime,t).
 \label{Eq. T int}
\end{equation}

By virtue of Eqs.~(\ref{Eq. G main})--(13), relation (\ref{Eq. T int}) can
be represented in the following form:

\begin{equation}
 T(\vec{r},t) \equiv T(z,t,\lambda) = T(z,t,\lambda=\infty) +\delta
 T(z,t,\lambda),
 \label{Eq. 23}
\end{equation}

\noindent  where  $T(z,t,\lambda=\infty)$  is the known solution of the
first boundary-value problem \cite{ref.1,ref.3}

\begin{equation}
 T(z,t,\lambda=\infty) = T_0 \left\{\sum_{n=0}^{\infty}
 (-1)^n\left[\erfc(nl-\xi) + \erfc\left((n+1)l+\xi\right)\right]-1\right\}
 \label{Eq. 24}
\end{equation}

\noindent and the additional term  $\delta T(z,t,\lambda)$  caused by
nonzero value of the quantity  $\lambda^{-1}$  has the form

\begin{eqnarray}
 \delta T(z,t,\lambda) &=& T_0 \left\{\sum_{n=1}^{\infty}
 \sum_{k=0}^{2n-1} C_{2n}^{k+1}\left[R_{k,n}(z)+R_{k,n}(L-z)\right]\right.
 \nonumber\\
 &+& \left. \sum_{n=0}^{\infty} \sum_{k=0}^{2n}
 C_{2n+1}^{k+1}\left[R_{k,n}(-z)+R_{k,n}(z-L)\right]\right\}.
 \label{Eq. 25}
\end{eqnarray}

\noindent  Here,  $l = L/(4\kappa t)^{1/2},$  $\xi = z/(4\kappa t)^{1/2},$
and

\begin{eqnarray}
 R_{k,n} &=& \frac{1}{\tilde{\lambda}\pi^{1/2}} \sum_{m=0}^k (-2)^{k-m-1}
 \left[\exp\left(-\left((2n+l)l-\xi\right)^2\right)
 \beta_m((2n+1)L-z,\lambda)\right.
 \nonumber\\
 &-& \left.\exp\left(-(2nl-\xi)^2\right)\beta_m(|2nL-z|,\lambda)\right].
 \label{Eq. 26}
\end{eqnarray}

\noindent  It is worth to note that expressions (23)--(25) can be written
at once by using Eq.~(72) in \cite{ref.8} for the density of Brownian
particles in a layer with absorbing boundaries and carrying out changes in
the corresponding notation in it.

As is seen from Eqs.~(23)--(26), distribution of the temperature in the
layer remains symmetric about the middle of the layer at any time  $t > 0:
\quad T(L-z,t,\lambda) = T(z,t,\lambda),$  which is quite natural due to
the symmetry of the system under investigation about the plane  $z = L/2$
and uniformity of the initial distribution of temperature.

By collecting terms with the same indices  $m$  in Eq.~(25), we can
represent the quantity  $\delta T(z,t,\lambda)$  as follows:

\begin{equation}
 \delta T(z,t,\lambda) = \frac{T_0}{\tilde{\lambda}\pi^{1/2}}
 \left\{\tilde{V_0}(z,\lambda) + \sum_{n=1}^\infty
 \left[V_n(z,\lambda)+\tilde{V_n}(z,\lambda)\right]\right\},
 \label{Eq. 27}
\end{equation}

\noindent where

\begin{eqnarray}
 V_n(z,\lambda) &=& \sum_{m=1}^{2n-1} l_{m,n}V_{m,n}(z,\lambda),  \qquad
 n=1,2,\ldots,
 \nonumber\\
 \tilde{V}_n(z,\lambda) &=& \sum_{m=0}^{2n} \tilde{l}_{m,n}
 \tilde{V}_{m,n}(z,\lambda),  \qquad n=0,1,2,\ldots,
 \nonumber\\
 V_{m,n}(z,\lambda) &=& \exp\left(-{\left(\xi_n^{(-)}\right)}^2\right)
 \eta_m(z_n^{(-)},\lambda) - \exp\left(-{\left(\xi_n^{(+)}\right)}^2\right)
 \eta_m(z_n^{(+)},\lambda)
 \nonumber\\
 &+& \exp\left(-{\left(\tilde{\xi}_n^{(+)}\right)}^2\right)
 \eta_m(\tilde{z}_n^{(+)},\lambda)
 - \exp\left(-{\left(\tilde{\xi}_{n+1}^{(-)}\right)}^2\right)
 \eta_m(\tilde{z}_{n+1}^{(-)},\lambda),
 \nonumber\\
 \tilde{V}_{m,n}(z,\lambda) &=& \exp\left(-{\left(\xi_n^{(+)}\right)}^2\right)
 \eta_m(z_n^{(+)},\lambda) - \exp\left(-{\left(\xi_{n+1}^{(-)}\right)}^2\right)
 \eta_m(z_{n+1}^{(-)},\lambda)
 \nonumber\\
 &+& \exp\left(-{\left(\tilde{\xi}_{n+1}^{(-)}\right)}^2\right)
 \eta_m(\tilde{z}_{n+1}^{(-)},\lambda)
 -\exp\left(-{\left(\tilde{\xi}_{n+1}^{(+)}\right)}^2\right)
 \eta_m(\tilde{z}_{n+1}^{(+)},\lambda),
 \nonumber\\
 \eta_n(z,\lambda) &=& (-1)^n 2^{-(n+1)}\beta_n(|z|,\lambda),  \qquad  n =
 0,1,2,\ldots,
 \label{Eq. 28}
\end{eqnarray}

\begin{eqnarray}
 z_n^{(\pm)} = 2nL \pm z, \quad  \tilde{z}_n^{(\pm)} = (2n-1)L \pm z,
 \quad  \xi_n^{(\pm)} = z_n^{(\pm)}/(4\kappa t)^{1/2},  \quad
 \tilde{\xi}_n^{(\pm)} = \tilde{z}_n^{(\pm)}/(4\kappa t)^{1/2}.
 \nonumber
\end{eqnarray}

The quantities  $l_{m,n}$  and  $\tilde{l}_{m,n}$  are defined by the
recurrence relations

\begin{eqnarray}
 l_{m,n} &=& l_{m-1,n} + (-1)^m 2^{m-1} C_{2n}^m,  \qquad  n=2,3,\ldots;
 \quad  m=1,2,\ldots,2n-1,
 \nonumber\\
 l_{1,n} &=& -2n,
 \nonumber\\
 \tilde{l}_{m,n} &=& \tilde{l}_{m-1,n} + (-1)^m 2^{m-1} C_{2n+1}^m,
 \qquad  n=1,2,\ldots;  \quad  m=1,2,\ldots,2n,
 \nonumber\\
 \tilde{l}_{0,n} &=& 1,  \qquad  \tilde{l}_{1,n} = -2n,
 \label{Eq. 29}
\end{eqnarray}

\noindent  which are convenient for  $m  \le  n.$  For  $m > n,$  the
following relations are more convenient:

\begin{eqnarray}
 l_{m,n} &=& l_{m+1,n} + (-2)^m C_{2n}^{2n-m-1},  \qquad  m < 2n-1,
 \nonumber\\
 l_{2n-1,n} &=& -2^{2n-1},
 \nonumber\\
 \tilde{l}_{m,n} &=& \tilde{l}_{m+1,n} + -(2)^m C_{2n+1}^{2n-m},
 \qquad  m \le 2n-1,
 \nonumber\\
 \tilde{l}_{2n,n} &=& 2^{2n}.
 \label{Eq. 30}
\end{eqnarray}

Relation (27) is valid for any  $\lambda > 0.$  In particular case of
great values of  $\tilde{\lambda},$  which corresponds holding the first
correction terms caused by a finite value of the relative heat transfer
coefficient of the layer with the environment, relation (27) is simplified
to the form

\begin{eqnarray}
 \delta T(z,t,\lambda) &=& \frac{T_0}{\tilde{\lambda}\pi^{1/2}}
 \left\{\sum_{n=-\infty}^\infty \frac{1-(1+|\gamma(n)|)(1+2|\gamma(n)|)^k}
 {\gamma(n)(1+|\gamma(n)|)} \exp\left(-(2nl-\xi)^2\right) \right.
 \nonumber\\
 &-& \left. \frac{1-(1+|\alpha(n)|)(1+2|\alpha(n)|)^m}
 {\alpha(n)(1+|\alpha(n)|)}
 \exp\left(-((2n+1)l-\xi)^2\right)\right\},
 \label{Eq. 31}
\end{eqnarray}

\noindent  where  $\alpha(n) = ((2n+1)l-\xi)/\tilde{\lambda},$  $\gamma(n)
= (2nl-\xi)/\tilde{\lambda},$  $n = 0,\pm1,\pm2,\ldots,$

\[
 m =\left\{ \begin{array}{rl}
  2n, & n \ge 0,\\
  -(2n+1), & n < 0,\\
  \end{array} \right.  \qquad  \qquad
 k = \left\{ \begin{array}{rl}
  2n-1, & n > 0,\\
  -2n, & n \le 0.\\
  \end{array} \right.
\]

For a thick layer ($l \gg 1$), relations (23), (24), and (31) for
distribution of the temperature can be simplified.  In this case, it is
sufficient to retain only terms with  $n = 0, 1$  in sum (24) and the term
with  $n = 0$ in sum (31).  This yields

\begin{equation}
 T(z, t,\lambda) \approx T^{sb}(z,t,\lambda) + T^{sb}(L-z,t,\lambda) -T_0,
 \label{Eq. 32}
\end{equation}

\noindent  where  $T^{sb}(z,t,\lambda)$  and   $T^{sb}(L-z,t,\lambda)$ are
the known distributions of the temperature in semibounded bodies
\cite{ref.7} that occupy the  $z \ge 0$  and  $z \le L$  regions,
respectively.  In the case  $\tilde{\lambda} \gg 1,$  they have the form

\begin{equation}
 T^{sb}(z,t,\lambda) \approx T_0 \left[\erf\xi +\frac{\exp(-\xi^2)}
 {\pi^{1/2}(\xi+\tilde{\lambda})}\right],
 \label{Eq. 33}
\end{equation}

\noindent  where

\[
 \erf\xi = 1 - \erfc \xi = \frac{2}{\pi^{1/2}}
 \int\limits_0^\xi dx\,\exp(-x^2).
\]

\newpage
\begin{figure}
 \centering{\epsfig{figure=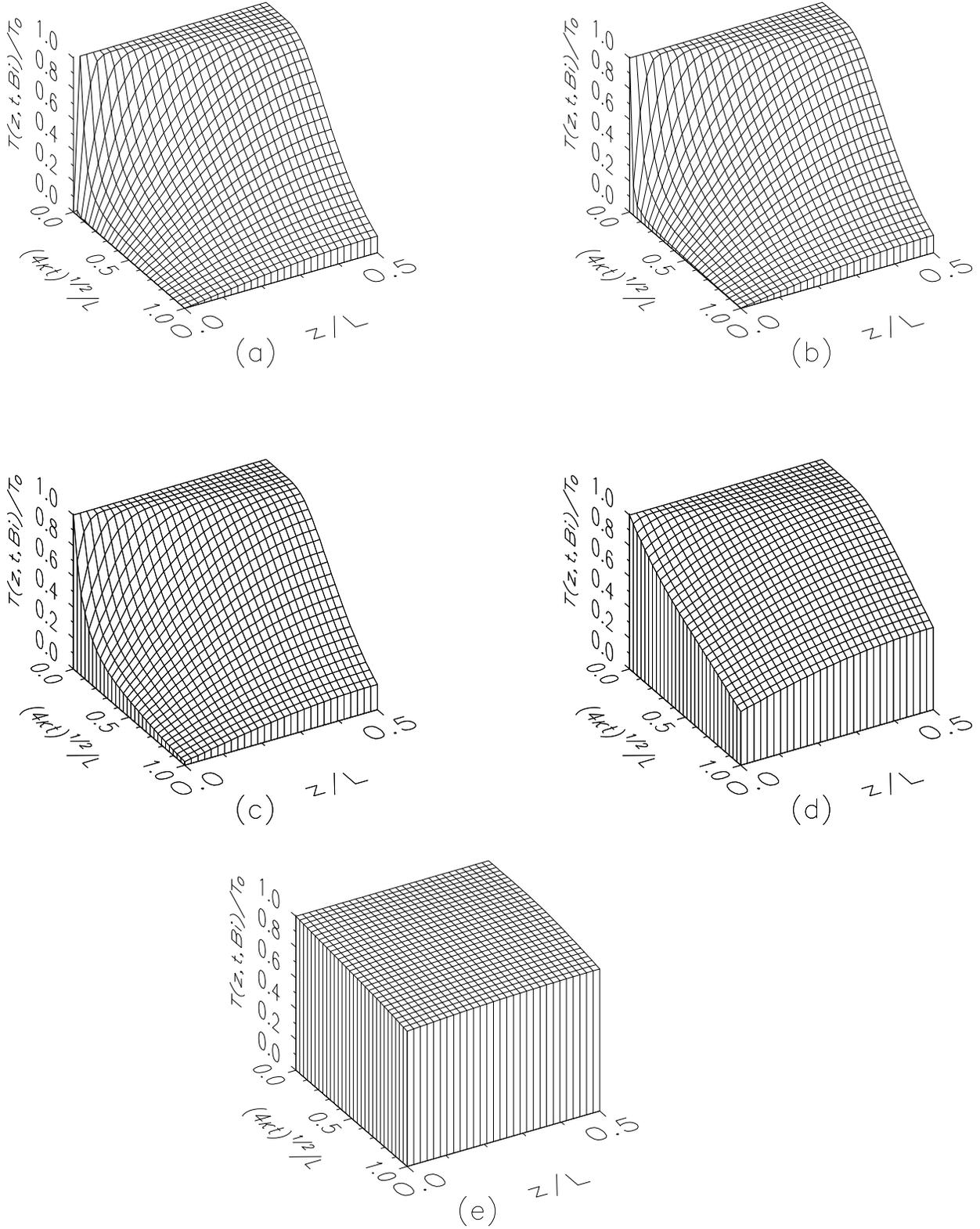,width=1.0\textwidth}}
 \medskip
 \caption{Space and time distribution of the temperature in the layer:
 $\Bi = \infty$ (a), 100 (b), 10 (c), 1 (d), and 0.1 (e).}
 \label{fig._2}
\end{figure}
\newpage

The space and time distribution of the normalized temperature
$T(z,t,\Bi)/T_0$  in the layer calculated by the general formulae
(23)--(24), (27)--(30) are displayed in Fig.~2 for various Biot numbers
$\Bi = L\lambda/2.$  With regard for the symmetry of the system at hand
about the  $z = L/2$  plane, the distribution of temperature is shown only
for the  $0 \le z \le L/2.$  It is quite natural that the layer cools down
with maximum rate in the case corresponding to the first-boundary problem
($\Bi = \infty$) when the boundaries of the layer are kept at zero
temperature for the entire time (Fig.~2a).  Note that a noninstanteneous
drop in the temperature shown in Fig.~2a from the initial value  $T_0$  to
zero at the initial time ($t = 0$) as the observation point approaches the
boundary of the layer is not the result of calculations but appears due to
the used uniformity of the space and time grid for the temperature field.
For finite great values of the Biot number (Fig.~2b), the distribution of
temperature inside the layer, in fact, does not change, whereas it takes
place the smooth but fast cooling of the boundary of the layer.  For the
time of order of  $0.1\tau,$  where  $\tau = L^2/4\kappa,$  the surface
temperature is as small as 6\% of its initial value.  As the Biot number
decreases ($\Bi =10,$  Fig. 2c), the cooling of the surface in noticeably
slows down and the surface temperature tends to zero even for  $t \approx
\tau.$  In this case, the finite value of $\lambda$  influences the
distribution of temperature deep into the layer, e.g.,
$T(L/2,\tau,\Bi=10)/T(L/2,\tau,\Bi=\infty) \approx 1.52.$  For
intermediate values of the Biot number (Fig. 2d), heat removal from the
surface does not lead to a radical redistribution of the temperature in
the central and peripheral regions of the layer as for  $\Bi \gg 1.$  For
example, for  $\Bi = 1$  and  $t = \tau,$  the ratio
$T(0,\tau,\Bi)/T(L/2,\tau,\Bi)$  is only 0.65.  For small Biot numbers
($\Bi = 0.1,$  Fig. 2d), the temperature equalization inside the layer
proceeds much faster than heat removal from the surface.  This case is
characterized by a slightly uniform distribution of temperature across the
section of the layer [$T(0,\tau,\Bi)/T(L/2,\tau,\Bi) \approx 0.95$] and a
slow decrease in the temperature with time [$T(L/2,\tau,\Bi)/T_0 \approx
0.92$].


\section{Conclusions}  \label{Concl.}

In the present paper, we gave a new representation for a solution of the
third boundary-value problem of the heat conduction equation for a layer
which needs no calculations of eigenvalues on the basis of the
transcendental equation.  This solution is represented in the form of a
superposition of the fundamental solutions for an unbounded system caused
by instantaneous heat sources situated at the points of location of
sources--images of the layer.  It is the form of representation of the
solution that is widely used in the literature for representation of
solutions of the first and second boundary-value problems.

Since new representation (9) enables one to carry out analytic
investigations not using graphic and table data, it can be useful for
various physical problems for which, in parallel with numerical
calculations, analytic estimations of solutions must be performed. In
particular, we can find the first additional terms caused by the account
of the heat transfer of the layer with the environment to the known
results determined under the assumption of a given temperature of the
surface or the absence of the thermal flow through it.  In the paper, we
illustrate the possibility of finding correction terms in powers of
$\lambda^{-1},$  where  $\lambda > 1,$  for a special problem of the
evolution of the initially uniform temperature field in the layer.

In addition to analytic investigations, Green's function (9) can be also
useful for the numerical analysis of various boundary-value problems of
the third order.  Using recurrence relations (14), we significantly
simplify this analysis by reducing it to the calculation of probability
integrals and the majorant estimation of the remainder terms of the series
given in the Appendix.

\newpage
\appendix
\section*{}

We prove that solution (9) satisfies the homogeneous heat conduction
equation (1) with boundary conditions (2) and (3).  For this purpose,
first, we prove the uniform convergence of the series in solution (9) at
$0 \le z \le L$  and  $t > 0.$  In view of the following representation of
the quantity  $\beta_k(|Z_\pm(n)|,\lambda):$

\begin{equation}
 \beta_k(|Z_\pm(n)|,\lambda) \approx (-1)^k \left(\frac{4\lambda\kappa t}
 {(|Z_\pm(n)| + 2\lambda\kappa t)}\right)^{k+1},
 \label{Ap. 1}
\end{equation}

\noindent  which is valid for  $|n| > (\kappa t)^{1/2}/L,$  we obtain that
for finite values of  $\lambda,$  for all  $|n| \ge N,$  where  $N \gg
\lambda\kappa t/L,$  the coefficients  $P_n^{(\pm)}(Z_\pm(n),\lambda)
\equiv P_n^{(\pm)}(\lambda)$  can be represented as follows:

\begin{eqnarray}
 P_n^{(-)}(\lambda) &\approx& \left(1 - \frac{2\lambda\kappa t}{L|n|}
 \right)^{2|n|},
 \nonumber\\
 P_n^{(+)}(\lambda) &\approx& \left(1 - \frac{2\lambda\kappa t}{L|n|}
 \right)^{2|n|+\sign n}.
 \label{Ap. 2}
\end{eqnarray}

\noindent  As  $|n| \to \infty,$  the following estimate for the
coefficients holds:

\begin{equation}
 \lim_{|n| \to \infty} P_n^{(\pm)}(\lambda) =
 \exp\left(-\frac{4\lambda\kappa t}{L}\right) \le 1.
 \label{Ap. 3}
\end{equation}

\noindent  Thus, starting from  $|n| \ge N,$  the series
$\sum\limits_{|n|=N}^\infty P_n^{(\pm})\exp\left(-Z_\pm^2(n)/4\kappa
t\right)$  in solution (9) can be majorized by the uniformly convergent
series  $\sum\limits_{|n|=N}^\infty\exp\left(-Z_\pm^2(n)/4\kappa
t\right).$  If follows from here that series in solution (9) are also
uniformly convergent.  Therefore, we can integrate solution (9) term by
term and use the majorant estimate

\begin{eqnarray}
 R(z,z^\prime,t) &=& \frac{1}{(4\pi\kappa t)^{1/2}} \left|
 \sum_{|n|=N}^{\infty} \left [P_n^{(-)}(\lambda)
 \exp\left(-\frac{Z^2_-(n)}{4\kappa t}\right) + P_n^{(+})(\lambda)
 \exp\left (-\frac{Z^2_+(n)}{4\kappa t}\right)\right]\right|
 \nonumber\\
 &\le& \frac{1}{l} \erfc \frac{(2N-1)L}{(\kappa t)^{1/2}}
 \label{Ap. 4}
\end{eqnarray}

\noindent for the remainder of the series which determines Green's
function (9).

It is easy to show that

\[
 \lim_{|n| \to \infty} \frac{\partial P_n^{(\pm)}(Z_\pm(n),\lambda)}
 {\partial z} = 0.
\]

Therefore, the series of derivatives

\[
 \sum_{n=-\infty}^\infty \left [\tilde{P}_n^{(-)}(Z_-(n),\lambda)
 \exp\left(-\frac{Z^2_-(n)}{4\kappa t}\right) +
 \tilde{P}_n^{(+)}(Z_n^{(+)},\lambda)\exp\left(-\frac{Z^2_+(n)}
 {4\kappa t}\right)\right],
\]

\noindent  where

\[
 \tilde{P}_n^{(\pm)}(Z_\pm(n),\lambda) = \frac{\partial P_n^{(\pm)}
 (Z_\pm(n),\lambda)}{\partial z}  - \frac{Z_\pm(n)}{2\kappa t}
 P_n^{(\pm)}(Z_\pm(n),\lambda),
\]

\noindent  converges uniformly as well because the series
$\sum\limits_{|n|=N}^{\infty} \tilde{P}_n^{(\pm)}(Z_\pm(n),\lambda)
\exp\left(-Z_\pm^2(n)/4\kappa t\right)$  are majorized by the uniformly
convergent series  $-\sum\limits_{|n|=N}^{\infty} \left(nL/\kappa t\right)
\exp\left(-Z_\pm^2(n)/4\kappa t\right).$  Thus, solution (9) is
term-by-term differentiable with respect to  $z.$  The possibility of
double term-by-term differentiation with respect to  $z$  and
differentiation with respect to  $t$  is proved in a similar way.

We represent Green's function (9) in the form

\begin{equation}
 G(\vec{r},\vec{r}{\,}^\prime,t) = \sum_{\alpha=\pm} \sum_{n=-\infty}^{\infty}
 G_n^{(\alpha)}(\vec{r},\vec{r}{\,}^\prime,t),
 \label{Ap. 5}
\end{equation}

\noindent  where

\begin{equation}
 G_n^{(\pm)}(\vec{r},\vec{r}{\,}^\prime,t) = P_n^{(\pm)}(Z_\pm(n),\lambda)
 \frac{\exp{\left(-\frac{R_\perp^2+Z_\pm^2(n)}{4\kappa t}\right)}}
 {(4\pi\kappa t)^{3/2}},  \qquad  n = 0, \pm1,\pm2,\ldots,
 \label{Ap. 6}
\end{equation}

\noindent  substitute expression (A5) into homogeneous equation (1), and
carry out term-by-term differentiation of the series.  By using recurrence
relation (14) for the functions  $F_k(x)$,  we obtain that expression (A5)
satisfies the required equations, moreover, each term
$G_n^{(\alpha)}(\vec{r},\vec{r}{\,}^\prime,t)$  of series (A5) satisfies
this equation.  By substituting solution (A5) into initial (6) and
boundary (2) and (3) conditions, we make sure that the solution satisfies
these conditions.  Note that, as opposed to the heat conduction equation,
boundary conditions (2) and (3) are valid not separately for each term
$G_n^{(\alpha)}(\vec{r},\vec{r}{\,}^\prime,t)$  of series (A5) but for
pairs, namely:

\begin{eqnarray}
 \left(\frac{\partial}{\partial z} - \lambda\right)\left(
 G_n^{(-)}(\vec{r},\vec{r}{\,}^\prime,t) + G_{-n}^{(+)}(\vec{r},\vec{r}{\,}^\prime,
 t)\right) &=& 0 \biggr |_{z=0},  \qquad  n = 0,\pm1,\pm2,\ldots,
 \nonumber\\
 \left(\frac{\partial}{\partial z} + \lambda\right)\left(
 G_n^{(-)}(\vec{r},\vec{r}{\,}^\prime,t) + G_{-(n+1)}^{(+)}(\vec{r},\vec{r}{\,}^\prime,t)
 \right) &=& 0 \biggr |_{z=L},  \qquad  n = 0,\pm1,\pm2,\ldots.
 \nonumber
 \end{eqnarray}

Thus, relation (9) is the solution of the third boundary-value problem of
the heat conduction equation.

\newpage


\end{document}